# Spectral and temporal modulations of femtosecond SPP wave packets induced by resonant transmission/reflection interactions with metal-insulator-metal nanocavities

NAOKI ICHIJI,[1] YUKA OTAKE,[1] ATSUSHI KUBO,[2,*]

[1]*Graduate School of Pure and Applied Sciences, University of Tsukuba, 1-1-1 Tennodai, Tsukuba-shi, Ibaraki 305-8573, Japan*
[2]*Division of Physics, Faculty of Pure and Applied Sciences, University of Tsukuba,1-1-1 Tennodai, Tsukuba-shi, Ibaraki 305-8573, Japan*
*\*kubo.atsushi.ka@u.tsukuba.ac.jp*

**Abstract:** To study the dynamical optical interactions of nano-scaled metal-insulator-metal (MIM) structures in temporal-frequency domain, femtosecond surface plasmon polariton (SPP) wave packets propagate over a surface with a MIM structure. The resonance nature of the SPP-cavity interaction is reflected as strong modulations in the spectra of transmitted and reflected SPP wavepackets, which show peaks and valleys, respectively, corresponding to the MIM cavity's eigenmode. These features indicate that the MIM structure acts as a Fabry–Pérot etalon-type spectrum filter. With appropriate tuning of the resonance frequency of the cavity, one can extract a wave packet with a narrower time duration and temporally shifted intensity peak.

## 1. Introduction

Surface plasmon polaritons (SPPs) are electromagnetic waves that represent the collective oscillation of free electrons confined at interfaces between metals and insulators [1, 2]. Because of the Drude responses of metals to external light fields [3], irradiations of laser pulses on metal surfaces equipped with a coupling structure such as a nano-scaled groove or ridge, launch SPPs in which the carrier wave replicates the frequency and phase of the excitation laser field [4-6]. Therefore, when an ultrashort laser pulse with a femtosecond duration is used, the excited SPP shapes a femtosecond wave packet (WP) with a time duration comparable to the excitation pulse [7-10]. SPP WPs propagate over metal surfaces or along artificially designed waveguide structures for 10 to hundreds of femtoseconds, and if they reach a nano-scaled object within the coherence lifetime, WPs cause optical excitations in the object [11-13]. This scenario constitutes the working principle of many plasmonic signal processing devices [14, 15]. Because successions of SPP pulses carry information in plasmonic devices, understanding the optical interactions between ultrashort SPP WPs and nano-scaled objects is crucial for designing wideband, high-speed plasmonic communication devices [16, 17].

This paper presents our study of optical interactions between femtosecond SPP WPs and nano-scaled plasmonic cavities consisting of a metal-insulator-metal (MIM) laminar structure fabricated on a flat metal surface [18]. The MIM structure, a well-studied SPP waveguide, is constructed by sandwiching a thin insulator layer between two metal layers [19, 20]. The structure supports transverse magnetic modes in which the field distributions are tightly confined in the insulator layer and the wavelengths are squeezed much shorter than light wavelengths in a vacuum [21, 22]. MIM nano-cavities are prepared by cutting the length of a MIM waveguide to nanometer scales [23, 24]. The supported SPP waves are reflected at both ends, and therefore the MIM nano-cavity can function as a Fabry–Pérot resonator when the length coincides with multiples of half the wavelength of the eigenmode [25-28]. Because of the significantly shortened eigenmode wavelengths, the MIM nano-cavity dimensions can be far below the sub-wavelength range. MIM nano-cavities interact strongly with external light fields, and moreover, they support magnetic resonances at visible frequencies because of the polarization current at the two interfaces between the insulator and metals [29-31]. For this reason, MIM nano-cavities are widely used as a meta-atom to construct two-dimensional meta-materials and meta-surfaces [32, 33]. The outstanding manufacturability due to their simple structure and their interesting properties, typified by their negative permittivity [34], make MIM nano-cavities appealing for applications in a wide range of fields. A number of such applications, including a negative refraction lens [35], Bragg reflectors [36, 37], perfect absorber [38-40], and anomalous reflection meta-surface [41, 42], have been studied theoretically and experimentally.

Particularly, combinations of femtosecond light pulses and meta-materials have enabled experimental verifications of anomalous group and phase velocities of light in materials having negative index of refraction [43]. The negative permittivity, which is the requirement for the negative index of refraction, arises as the dispersive property near the resonance frequency of meta-atoms. The frequency-dependent modulations in both the phase and the amplitude of light fields are accumulated over the spectral bandwidth and the sectional area of the incident light beam to determine the entire shape of the temporal waveform and the spatial distribution of the wave front of the light pulses. Therefore, investigations of the dynamical responses of individual meta-atoms are indispensable for understanding the anomalous light propagations in meta-materials. Numerical simulations based on finite-difference time-domain (FDTD) method can be a powerful tool for this purpose, because of its capability for tracking time-dependent dynamics of light propagations and interactions in materials [44]. However, most numerical analyses have been more focused on static optical properties for a specific frequency range, and only a few reports dynamics of light or SPP WPs interact with meta-atoms.

In this study, single MIM nano-cavities were chosen as the site of interaction with SPP WPs. We report FDTD simulations that reveal spectral and temporal modulations of femtosecond SPP WPs induced by resonant transmission and reflection interactions with the MIM nano-cavities. Time-frequency analyses of temporal waveforms of the transmitted and the reflected SPP WPs elucidate a complemental relation between these two wave components is conserved in both the spectral and the time domains. Interestingly, when the incident SPP WP is chirped, MIM nano-cavities with the range of the length of 150-200 nm emit a SPP WP with a constricted time duration as the transmission wave, and another dark SPP pulse incised in a WP as the reflection wave. The intensity peak of the transmitted SPP WP can be tuned temporally for either the forward or the retard directions in the range of several femtoseconds. Section 2 describes the models used for the FDTD simulations and the analysis of MIM nano-cavity eigenmodes. Section 3 discusses the time-frequency analysis of transmitted and reflected WPs by nano-blocks and MIM nano-cavities. The results demonstrate that the MIM nano-cavity acts as a spectrum filter. Section 4 details the effects of the MIM nano-cavities on incident SPP wavepackets by comparing the spectra and waveforms of the transmission and reflection waves, and Section 5 presents the conclusions of this study.

## 2. Structure and optical properties of MIM nano-cavity

### 2.1 Setup of FDTD simulation

Figure 1 (a) shows a schematic of the MIM nano-cavity studied in this work. The bottom layer is composed of titanium (Ti), which is commonly used as an adhesive layer for depositing gold (Au) films on substrates. The thickness of the Ti film, $d = 500$nm, was chosen thick enough to eliminate any material-dependent effects from substrates. A comparison between a thinner Ti film ($d = 50$ nm) didn't show differences in major results of FDTD simulations. A Au film with a thickness $h$ of 100 nm and an $Al_2O_3$ film with a thickness $S$ of 16 nm were sequentially placed on the Ti layer. The MIM nano-cavities were formed by putting a Au nano-block with a thickness $h$ of 100 nm and a length $L$ ranging from 10 to 500 nm on the $Al_2O_3$ film. In addition, the whole surface was covered with a poly(methyl methacrylate) (PMMA) layer with a thickness $w$ of 60 nm to simulate experimental microscopic imaging conditions [45]. The only effect of this additional PMMA layer was a slight increase in the effective refractive index of the SPPs. The relative dielectric constants of Au and Ti were determined from Rakic *et al.* [46], with one Drude term and four Lorentz terms in the Lorentz-Drude model. Compared to the Drude model, the Lorentz-Drude model reproduces dielectric constants more accurately, particularly for both the real and imaginary components of the visible light region of Au, which is essential to prevent overestimations of SPP lifetimes. The relative dielectric constants of PMMA ($\epsilon_{PMMA}$ = 2.34) and $Al_2O_3$ ($\epsilon_{Al_2O_3}$ = 2.75) were set as non-dispersive constants. In addition to MIM nano-cavities (Fig. 1a), simple nano-blocks (Fig. 1b) and a flat surface (Fig. 1c) were also prepared to clarify how the resonant property of the MIM nano-cavity affects optical interactions with SPPs. Nano-blocks with a thickness $h'$ of 216 nm were formed by replacing the $Al_2O_3$ part of the MIM structures with Au.

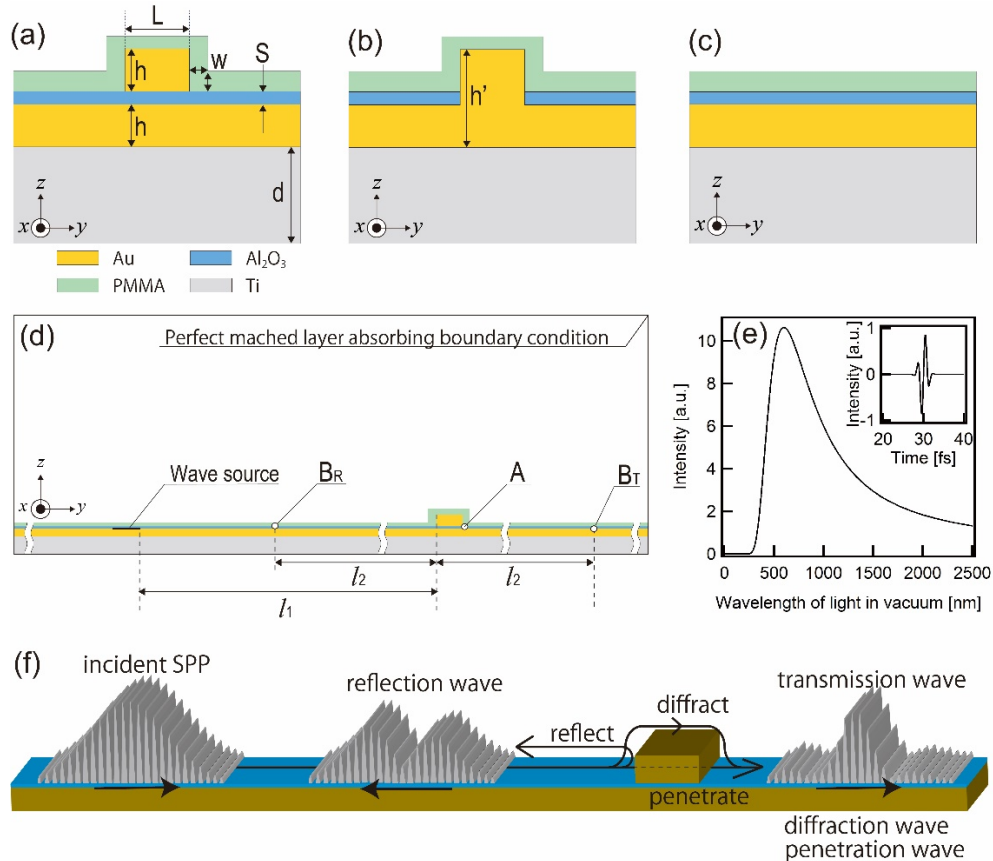

Fig.1 Schematics of multi-layered structures used for FDTD simulations: (a) MIM nano-cavity, (b) nano-block, (c) flat surface, and (d) the whole calculation area. (e) Spectrum of the excitation wave. Inset shows the waveform applied to the wave source. (f) A schematic model showing the propagation of a SPP WP and the interaction with a nano-structure, either a MIM nano-cavity or a nano-block. The SPP WP splits into reflection and transmission waves when it reaches the nano-structure. The transmission wave includes diffraction and penetration components.

Figure 1(d) shows the whole area of the FDTD simulation. Commercial FDTD software, Poynting for Optics FUJITSU, was used for all simulations. A periodic boundary condition was applied for the $x$-axis, and absorbing boundary conditions (perfect matched layers) were applied to both ends of the $y$- and $z$-axes. The length of the area along the $y$-direction was selected to be long enough to prevent artifacts caused by SPP waves reflected from the boundaries. We set the origin of the $y$-axis at the left edge of the MIM nano-cavity. A wave source for launching SPP WPs was placed at the $Al_2O_3$/Au boundary at $y$ = -4 μm ($l_1$ = 4 μm). Time evolutions of the vertical component of the electric fields ($E_z(t)$) at the center of the $Al_2O_3$ layer, *i.e.*, 8 nm above the Au film, were re-coded at $y = -3$ μm ($B_R$), $L$ ($A$), and 3 μm ($B_T$). The point $A$ corresponds to the right edge of the MIM nano-cavity.

To investigate broadband spectral features of MIM nano-cavities over the visible to near-infrared range, as well as to extract the femtosecond temporal behavior of SPP WPs, a very short oscillation field was applied to the wave source. The function of the waveform, shown in the inset of Fig. 1(e), was expressed as,

$$f(t) = \exp\left(-\left(\frac{t-t_0}{\Delta\tau}\right)^2\right)\sin(\Omega t), \tag{1}$$

where, $\Delta\tau$ was 1.2 fs, $t_0$ was 30 fs, and $\Omega$ was π rad/fs. The spectrum of the wave source is shown in Fig. 1(e). As shown in Fig. 1(f), the SPP WP launched from the wave source propagates on the Au surface, and when it reaches the MIM nano-cavity or nano-block, it divides into a transmission wave and a reflection wave, as well as a scattered wave. Here, we focus on the transmission and reflection components because the scattered wave no longer remains on the surface. We further divide the transmission wave into two components, analogous to the modeling of surface waves on nano-holed metal films [47]: one component is a SPP wave that overcomes the nano-structures, named the diffraction wave, and the other is a wave, called the penetration wave, that passes through the structures, whereby the MIM nano-cavity acts as a waveguide or the nano-block extinguishes the wave. As discussed later, the diffraction and penetration components occupy different frequency regions, and the latter component as well as the reflection wave reveal resonance characteristics of the nano-cavities.

## 2.2 Resonant field enhancement in MIM nano-cavity: FDTD and mode analysis

When the SPP WP reaches the MIM nano-cavity, the cavity's eigenmode is excited and the inner electromagnetic field is enhanced. We evaluated the resonant spectrum of the MIM nano-cavity, $R(\omega)$, as

$$R(\omega) = |\tilde{\mathcal{F}}_{\text{cav}}(\omega)|^2 / |\tilde{\mathcal{F}}_{ref}(\omega)|^2, \tag{2}$$

where $\tilde{\mathcal{F}}_{\text{cav}}(\omega)$ and $\tilde{\mathcal{F}}_{\text{ref}}(\omega)$ were, respectively, the fast Fourier transform (FFT) of the $E_z(t)$ field measured at the point $A$ at the end of cavity (Fig.1a) and at the center of the Al$_2$O$_3$ layer (Fig.1c).

Figure 2(a) shows the resonance spectrum obtained for a MIM nano-cavity with a length $L$ of 200 nm. The magnitude of the vertical axis indicates the degree of field enhancement. One can clearly see two resonance peaks at 1800 nm and 950 nm, which respectively correspond to the 1st and 2nd Fabry-Pérot resonant modes of the cavity [24]. Resonance modes can be described by the following equation:

$$Lk_0 n_{\text{eff}} + \phi = N\pi, \tag{3}$$

where $k_0$ is the vacuum wave number, $n_{\text{eff}}$ is the real component of the effective refractive index of the MIM nano-cavity, $N$ is the integer defining the order of the mode, and $\phi$ is the additional phase that accounts for the fact that eigen-functions of MIM nano-cavities are not immediately reflected from the physical boundaries of the block configuration [28]. From Eq. (3), the resonance frequency is rigorously connected to the cavity length $L$, and can thus be tuned by controlling the value of $L$. Figure 2(b) shows a series of resonance spectra calculated for MIM nano-cavities with different values of $L$, from 50 to 500 nm, in increments of 50 nm. As the cavity becomes longer, the wavelength of each resonance peak gets longer; the separation between adjacent resonant modes narrows; and eventually, more resonant modes appear in the limited wavelength range. This result suggests that a broadband SPP WP could excite multiple resonance modes, including both even and odd orders.

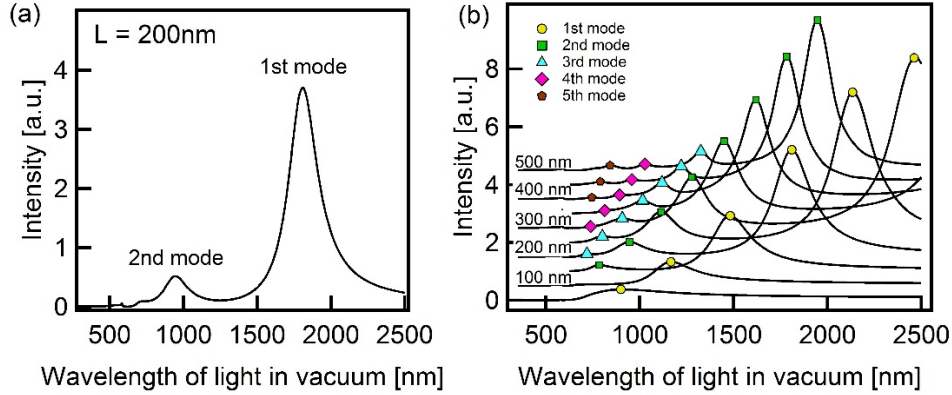

Fig. 2. (a) Resonance spectra of a MIM nano-cavity with a length $L$ of 200 nm. (b) A series of resonance spectra for the range of $L$ from 50 to 500 nm, at intervals of 50 nm. Each spectrum is offset by 0.5 along the vertical axis. The yellow, green, blue, pink, and brown marks correspond to the 1st, 2nd, 3rd, 4th, and 5th-order resonance peaks, respectively.

SPP waves propagate inside the MIM nano-cavity as through a waveguide. The MIM nano-cavity studied here was established by setting a Au block with a limited length on a continuous $Al_2O_3$/Au film (Fig. 1a); therefore, the geometry was asymmetric with respect to the center of the insulator layer. Nevertheless, this system is still useful for comparing resonance frequencies with analytically derived eigenfrequencies for a symmetric MIM waveguide.

The dispersion relation of the transverse-magnetic-polarized waveguide mode of the MIM structure configured as shown in Fig. 3(a) was derived in the following manner [24]. We focus on the so-called odd mode, where the spatial distribution of $E_y$ ($E_z$) is antisymmetric (symmetric) with respect to the center ($z = 0$) of the MIM structure. The length of the structure, $L$, is assumed to be infinite. Wave functions of the electromagnetic field that propagate in the $y$-direction from the wave vector $k$ and the frequency $\omega$ are then expressed as follows:

In $|z| > T/2$ (in the metal layer),

$$H_x^m(r, \omega) = A \frac{\omega \epsilon_0 \epsilon_m(\omega)}{k} \exp(-k_m|z|)\exp[i(ky - \omega t)], \tag{4}$$

$$E_y^m(r, \omega) = -\frac{ik_m}{k}\frac{|z|}{z} A \exp(-k_m|z|)\exp[i(ky - \omega t)], \tag{5}$$

$$E_z^m(r, \omega) = A \exp(-k_m|z|) \cdot \exp[i(ky - \omega t)], \tag{6}$$

and in $|z| < T/2$ (in the insulator layer),

$$H_x^i(r, \omega) = 2B \frac{\omega \epsilon_0 \epsilon_i}{k} \cosh(k_i z) \exp[i(ky - \omega t)], \tag{7}$$

$$E_y^i(r, \omega) = i2B \frac{k_i}{k} \sinh(k_i z) \exp[i(ky - \omega t)], \tag{8}$$

$$E_z^i(r, \omega) = 2B \cosh(k_i z) \exp[i(ky - \omega t)], \tag{9}$$

where, $\epsilon_0$ is the dielectric constant of the vacuum; $\epsilon_m(\omega)$ and $\epsilon_i$ are the relative dielectric constants of the metal (Au) and the insulator (Al$_2$O$_3$), respectively; A and B are arbitrary constants; sub- and super-script $m$ and $i$ indicate the metal and the insulator, respectively; and $k_m$ and $k_i$ are $z$-components of the wave vector, respectively written as $k_m = \sqrt{k^2 - \frac{\omega^2}{c^2}\epsilon_m(\omega)}$ and $k_i = \sqrt{k^2 - \frac{\omega^2}{c^2}\epsilon_i(\omega)}$. Applying boundary conditions for connecting Eqs. (4) and (7) and Eqs. (6) and (9) at $z = T/2$ results in

$$\epsilon_m(\omega) A \exp(-k_m|x|) = 2B\epsilon_i \cosh(k_i x), \tag{10}$$

and

$$-k_m A \exp(-k_m|x|) = 2B k_i \sinh(k_i x). \tag{11}$$

The eigenequation of the MIM structure is derived from Eqs. (10) and (11) as follows:

$$-\frac{k_m}{k_i} = \frac{\epsilon_m(\omega)}{\epsilon_i} \tanh\left(k_i \frac{t}{2}\right). \tag{12}$$

The relation of the wave vector $k$, or the wavelength $\lambda_{\text{MIM}} = 2\pi/k$ to the frequency $\omega$, i.e., the dispersion relation of the SPP in the MIM structure, is obtained by numerically solving Eq. (12). Assuming $\phi = 0$ in Eq. (3), the condition for the $N$-th order Fabry-Pérot resonance can then be rewritten as

$$k(\omega) = 2\pi \cdot \frac{N}{2L}, \tag{13}$$

which determines the relation between the cavity length and the frequency for the $N$-th mode.

The solid lines in Fig. 3(b) show the calculation results using Eq. (13) for $N = 1$–5, providing the relation between the light wavelength and the length of the cavity that supports resonances. These analytical results show reasonable agreement with the resonance wavelengths of the MIM nano-cavities determined from the peak intensities of resonance spectra in FDTD simulations (Fig. 3b, marked points), some of which are indicated in Fig. 2(b). Compared to the analytical solutions for the symmetrical MIM structure, the lengths of the cavities at resonances calculated by FDTD tended to shift to shorter lengths by approximately 60 nm. The origin of these shifts is the additional phase ($\phi$), as shown in Eq (3). The magnitude of $\phi$ varies according to the specific configuration, including the symmetry of the MIM structure. Pors and Bozhevolnyi reported that an asymmetric MIM structure constructed by a finite length metal strip resting on an infinite length insulator and metal layers shows strong redshift of the resonance compared to a symmetric MIM [18]. Our result were consistent with the previous report, as shown as Fig. 3(b).

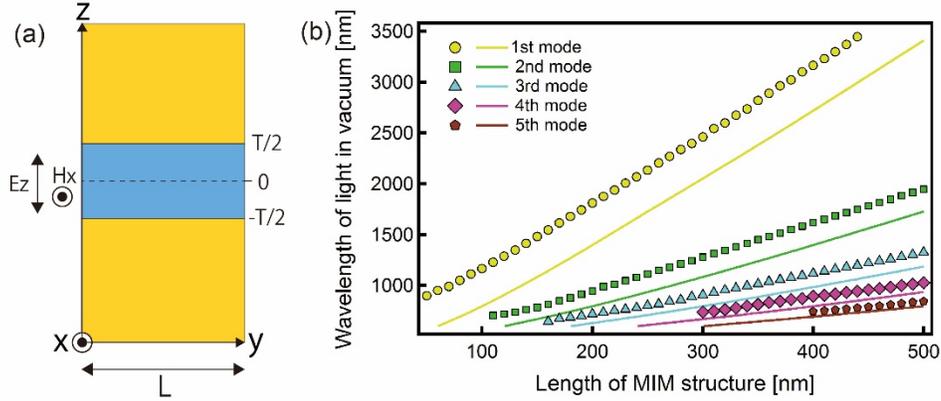

Fig. 3. (a) Schematic of MIM structure used for the analysis of the waveguide mode. The length of the structure along the *y*-axis is $L$, and the thickness of the insulator is $T$. The structure is uniform in the *x*-direction. (b) Relation between the length of a MIM structure and wavelengths of light that excite the structure's resonance modes (1st–5th order). Calculation results obtained by FDTD simulation (marked points) and by mode analysis using Eq. (13) (solid lines) are shown.

## 3. Time-frequency analyses of transmitted/reflected SPP WPs

### 3.1 Time-frequency spectrogram of SPP WPs

The temporal width of SPP WPs spreads significantly during the propagation over the Au surface because of the dispersive character of the SPP mode's complex dispersion relation: the curvature of the real component of the dispersion relation determines the group velocity dispersion of the SPP WP, and thus the rate of the temporal width expansion. The real component of the dispersion relation also limits the frequency range of SPP WPs from the red to near-infrared region according to the boundary condition, $\epsilon_{\text{Au}}(\omega) < -|\epsilon_i^{\text{effec}}|$, where $\epsilon_{\text{Au}}(\omega)$ and $\epsilon_i^{\text{effec}}$ are, respectively, the relative dielectric constant of Au and the effective relative dielectric constant of the vacuum/PMMA/$Al_2O_3$ layers. The imaginary component of the dispersion relation induces a narrowing of the spectral width and a redshift of the center frequency of SPP WPs.

Because the dispersion curve exhibits a normal dispersion over the red to near-infrared region, the carrier frequency of the broadened SPP WP is up-chirped, *i.e.*, the instantaneous frequency is low at the front of the SPP WP and gradually increases as time elapses. Therefore, the resonant interaction between a SPP WP and a MIM nano-cavity occurs at the specific time when the instantaneous carrier frequency matches the resonant frequency. To identify this time and investigate frequency-dependent interaction dynamics, we transformed the $E_z(t)$ fields in a time-frequency analysis using the Wigner distribution function (WDF) [48].

Figure 4(a) shows a temporal waveform of the $E_z$ component of the SPP field (black line) measured at $B_T$ on the flat surface (Fig. 1c), and its envelope shape (blue line). The time duration (FWHM) of the envelope shape was approximately 34 fs, showing a considerable broadening compared to the waveform of the excitation source (Fig. 1e). The carrier signal was obviously up-chirped. Figure 4(b) shows a time-frequency spectrogram of the $E_z$ field. The up-chirping of the carrier frequency was reflected in the spectrogram as a stretched contour distributed from the lower left to the upper right. Compared to the spectrum of the excitation source, the frequency range narrowed, particularly at higher frequencies, because of the dispersion relation of the SPP. Nevertheless, the spectrogram profile shows an overall continuous, smoothly changing shape with no specific peaks or dips. This feature is consistent with the FFT spectrum of the $E_z$ field shown in Fig. 4(c).

In contrast to propagation over the flat surface, the shape of the SPP WP was significantly modulated after passing through a MIM nano-cavity. Figure 4(d-f) show the temporal waveform (d), the time-frequency spectrogram (e), and the FFT spectrum (f) measured at $B_T$ after the SPP WP passed through a MIM nano-cavity with a length $L$ of 150 nm. As shown in Fig. 4(f), only some of the spectrum components could transmit through the MIM nano-cavity: the FFT spectrum shows a main peak at 1.6 eV and a broad sub-peak at 1.1 eV. The temporal waveform and the spectrogram show that these two spectral components dominated different time regions. The SPP WP nonetheless maintained the feature of a femtosecond wave packet after the transmission, although the time and the instantaneous carrier frequency at the maximum intensity shifted compared to the SPP WP propagated over the flat surface.

The frequency-selective transmission shown in Fig. 4(d-f) was attributed to a filtering function of the Fabry–Pérot resonance of the MIM structures [49-51]. Indeed, the energy of the main peak at 1.6 eV in Fig. 4(f) corresponded to the 2nd mode frequency of a MIM nano-cavity of $L = 150$ nm (Fig. 2a). The energy of the main peak varied sensitively as a function of $L$, as is discussed in detail in Section 3.3. In contrast, the energy of the sub-peak was insensitive to $L$. Moreover, similar small peaks were also observed for interactions of SPP WPs with nano-blocks (Fig. 1b) as discussed in Section 3.2. We therefore attributed the sub-peak to a SPP component that overcame the nano-cavity as a diffraction wave. The properties of these wave components, as well as features of spectral filtering of MIM nano-cavities, were reflected in the intensity distributions of the transmitted and reflected WPs in the time-frequency domain.

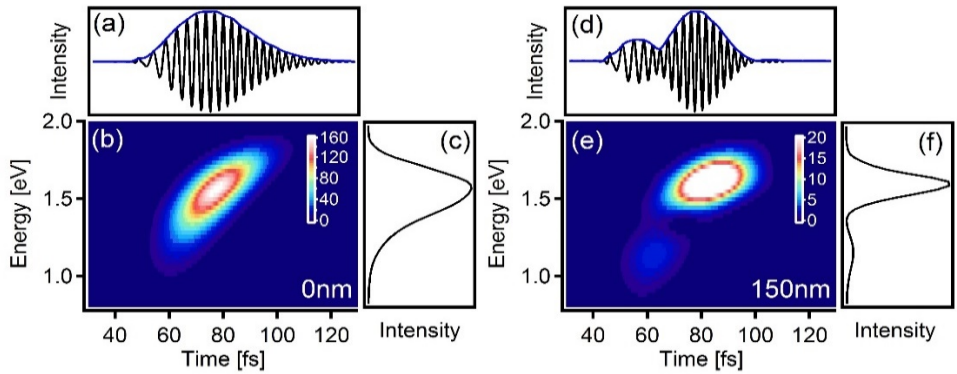

Fig. 4. (a) Temporal waveform (black line) and its envelope shape (blue line) of the $E_z$ component of a SPP WP measured at point $B_T$ of the flat surface. (b) Time-frequency spectrogram of (a). The color scale represents the intensity. (c) FFT spectrum of (a). (d-f) Same plots as shown in (a-c) for a SPP WP passed through a MIM nano-cavity with a length of 150 nm.

### 3.2 Transmission and reflection of SPP WPs: nano-blocks

We first investigated the transmission and reflection of SPP WPs for the Au nano-blocks (Fig. 1b). Figure 5(a) shows a series of envelopes of SPP WPs measured at $B_T$, *i.e.*, transmitted waves. The length of the nano-block varied from 0 to 250 nm in increments of 10 nm. (The length of 0 nm corresponds to the flat surface.) The intensity maximum of each envelope is indicated by a red circle in the figure. These marks show two peak locations: one at 65 fs and the other at 80 fs. Although the intensity of the latter peak decreased with increasing block lengths, that of the former peak remained at an almost constant value. Spectrograms of some typical block lengths are presented in Fig. 5(b-e) to show how these two components were distributed and how their intensities changed in the time-frequency space as the block length increased.

The latter component, of which the energy was distributed at approximately 1.6 eV, occupied a higher energy level than the former component, at approximately 1.1 eV, showing consistency with the energy distribution of the up-chirped carrier wave of the incident SPP WP. Figure 5(f) shows the peak intensity of the latter component, at 1.6 eV, plotted as a function of the block length. The intensity damping trend was well-fitted by an exponential function. The value of the attenuation length, determined to be 30 nm, was in good agreement with the skin depth of gold at 1.6 eV [52]. This consistency suggests that the latter component, occupying the higher frequency region, penetrated inside the gold block to get across the nano-structure.

In contrast, the insensitivity of the former component's intensity to the block length suggested the existence of another pathway to circumvent the propagation inside of the block structure. Indeed, the height of the nano-blocks (100 nm) was significantly smaller than the wavelength of the SPP. Therefore, a portion of the SPP wave diffracted at the left edge of a nano-block could rise over the structure to reach the other side and still maintain the character of a surface wave. Therefore, transmitted SPP WPs can be regarded as the sum of a penetrated wave and a diffraction wave.

Although the waveform of transmitted SPP WPs changed as the block length changed, reflected WPs retained almost constant intensity and shape. Figure 6 shows a series of envelopes of reflected waves measured at $B_R$, and selected time-frequency spectrograms. Except for a gradual increase in the intensity maxima (red circles in Fig. 6a) as the block length increased from 10 to 50 nm, both the envelopes and spectrograms show almost identical waveforms. The envelope shape and the up-chirped carrier frequency were very similar to that of the SPP WP on a flat surface shown in Fig. 4(a-c). A FFT spectrum of the SPP WP reflected by the 250-nm-long nano-block is shown in Fig. 6(f) together with the spectrum on the flat surface (the reference wave) shown in Fig. 4(c). The two spectra had almost the same shape, suggesting nano-blocks equally reflected SPP frequency components that were distributed over the spectral width of the SPP WP. The reflectance ratio was estimated at approximately 0.6.

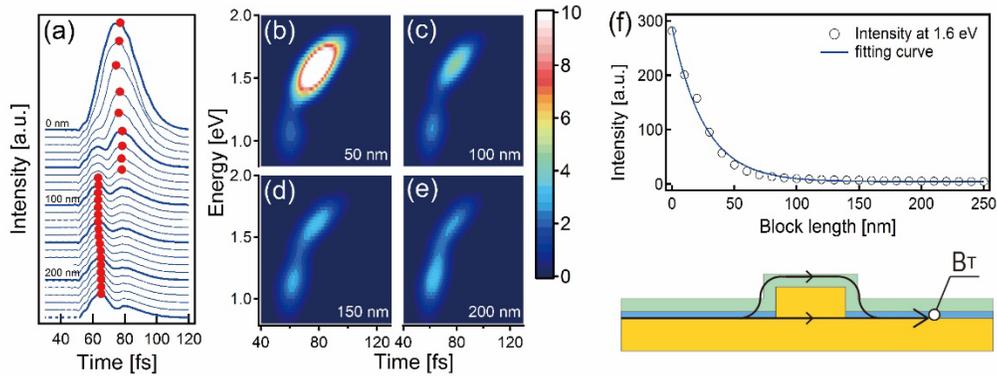

Fig. 5. (a) Envelope shapes of the $E_z$ component of transmitted SPP WPs measured at $B_T$ with a nano-block of length $L$ ranging from 0 to 250 nm. The envelopes are offset from one another; the lower envelopes correspond to longer $L$. Bold lines indicate multiples of $L = 50$ nm. Intensity maxima are indicated by red circles. (b-e) Time-frequency spectrograms of SPP WPs prepared for $L = 50, 100, 150,$ and 200 nm. Each frame corresponds to a bold line in (a). (f) Intensities of the transmitted component measured at 1.6 eV are plotted as a function of the block length $L$. The solid line shows the least-square fitting using an exponential function.

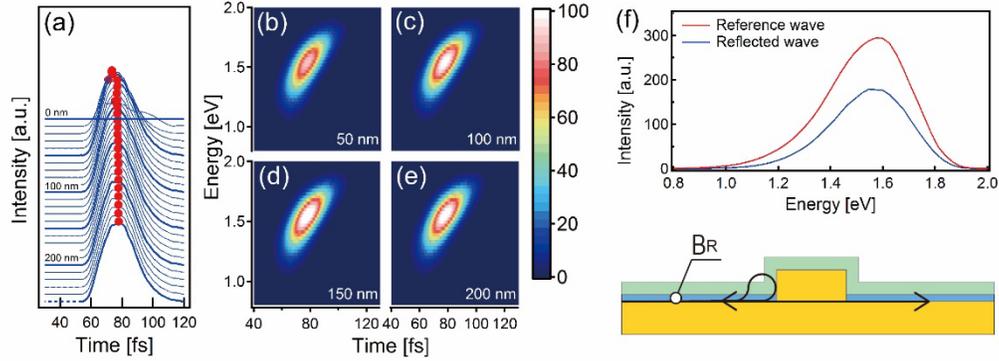

Fig. 6. (a) Envelope shapes of the $E_z$ component of reflected SPP wave packets measured at $B_R$ with placing a nano block of the length $L$ ranges from 0 to 250 nm. The envelopes are offset from one another; the lower envelopes correspond to longer $L$. Bold lines indicate multiples of $L = 50$ nm. Intensity maxima are indicated by red circles. (b-e) Time-frequency spectrograms of SPP WPs prepared for $L = 50, 100, 150,$ and $200$ nm. Each frame corresponds to a bold line in (a). (f) FFT spectrum of the reference SPP measured at $B_T$ of the flat surface, and the reflected SPP measured at $B_R$ for the block length of 250 nm.

## 3.3 Transmission and reflection of SPP WPs: MIM nano cavities

We next investigated the properties of the MIM nano-cavity by comparing the transmission and reflection components of SPP WPs. Envelope shapes of the transmitted SPP WPs measured at $B_T$ for cavity lengths of $L = 0$–500 nm are shown in Fig. 7(a). The intensity maximum of each envelope is indicated by a red circle. In contrast to the nano-block structure, the waveform changed in a more complex manner as $L$ increased: although all envelopes had almost the same shape before 60 fs, where the diffraction wave was considered to be the dominant component as in the block structure, in the range of 60–120 fs, there were multiple peaks. Moreover, the number of peaks and their appearance times varied as $L$ increased. In general, the appearance times of the intensity maxima tended to shift toward earlier times as $L$ elongated. Such trends can be seen for $L = 40$–100 nm, 110–430 nm, and 440–500 nm.

Because of the up-chirping of the carrier frequency of SPP WPs, peak shifts in the temporal domain correspond to peak shifts in the frequency domain. Figure 7(b-m) show time-frequency spectrograms of SPP WPs selected from the range of $L = 40$–480 nm at intervals of 40 nm. The sequence of spectrograms shows that the peak positions shift from the upper right to the lower left with increasing cavity length, indicating that the red shifts of the spectral components of transmitted SPP WPs are a function of $L$. This red shift of a peak associated with the temporal forwarding is clearly seen in Fig. 7(d-f). The peak intensities obtained with these nano-cavities were stronger than that for $L = 80$ nm (Fig. 7c) despite the extended length of the nano-structure, showing a contrasting effect compared to the nano-block. In Fig. 7(g), while the original peak continues to move toward the lower left, another peak appears at the upper right. These two peaks correspond to two adjacent peaks in the envelope shape of the nano-cavity with $L = 240$ nm indicated by the dashed lines in Fig. 7(a). In a similar manner as described above, both of these peaks again shifted to the lower left, and eventually merged into a single peak at the lower left corner (Fig. 7i). After a little while, as shown in Fig. 7(k), a new peak appeared at the upper right. These appearances of new peaks and their red shifts were related to the emergence of higher-order resonances and the peak shifts of the eigenmode of the MIM nano-cavity, as shown in Fig. 2(b), as detailed in Section 4.

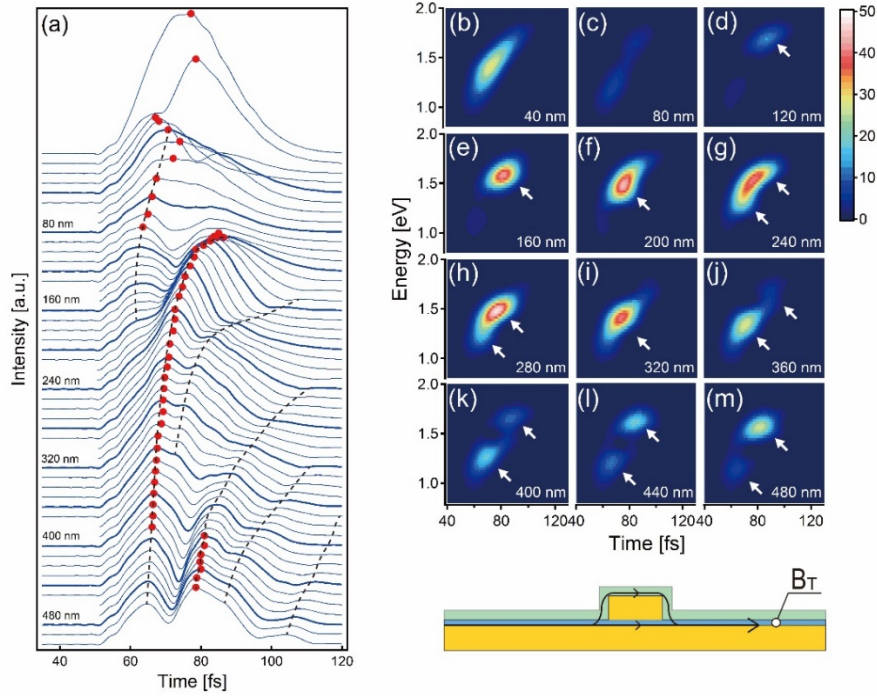

Fig. 7. (a) Envelope shapes of the $E_z$ component of transmitted SPP WPs measured at $B_T$ with a MIM nano-cavity of length $L$ ranging from 0 to 500 nm. The envelopes are offset from one another; the lower envelopes correspond to longer $L$. Bold lines indicate multiples of $L = 40$ nm. Intensity maxima are indicated by red circles. Dashed lines indicate positions of minor peaks. (b-m) Time-frequency spectrograms of SPP WPs corresponding to the bold lines in (a). White arrows indicate positions of intensity peaks.

Reflected SPP WPs showed complimentary behavior. Figure 8(a) and (b-m), respectively, show the envelope shapes of the reflected SPP WPs measured at $B_R$ and selected time-frequency spectrograms. For cavity lengths shorter than several tens of nanometers (Fig. 8b,c), spectrograms of the reflected WPs were rather similar to those obtained with the nano-blocks because the length $L$ was too short to support the cavity's eigenmodes. However, for cavity lengths longer than approximately 100 nm, eigenmodes of the MIM nano-cavity started to appear within the spectral range of the incident SPP WP, and both the envelope shapes and the spectrograms were largely affected by the resonances. In contrast to the transmission characteristics, the reflected effect was the generation of intensity "holes" in the spectrograms. A conspicuous example is Fig. 8(e), where a single hole appears at the energy of 1.55 eV and the time of 83 fs. These coordinates agree well with those of the peak of a transmitted SPP WP for the corresponding nano-cavity: the coordinates of the peak in Fig. 7(e) were 1.58 eV and 82 fs. As with the peaks in the transmitted SPP WPs, the holes in the reflected SPP WPs moved from the upper right to the lower left in the spectrogram as $L$ increased. Moreover, the number of holes increased, as indicated by the arrows in the spectrograms, again showing a correlation with the peaks of transmitted WPs.

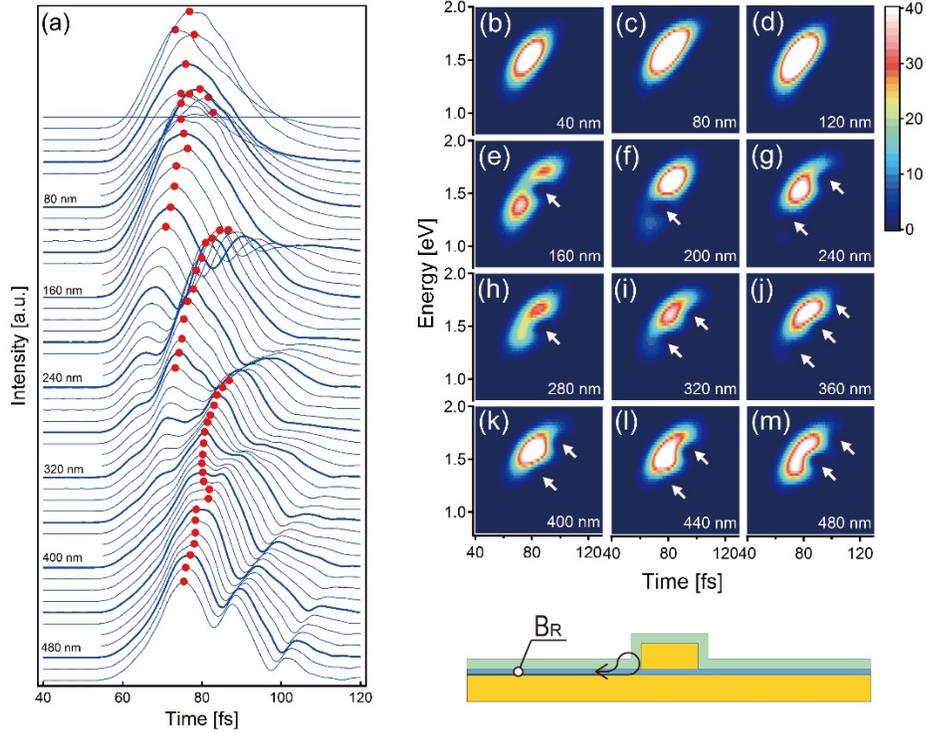

Fig. 8(a) Envelope shapes of the $E_z$ component of reflected SPP WPs measured at $B_R$ with a MIM nano-cavity of length $L$ ranging from 0 to 500 nm. The envelopes are offset from one another; the lower envelopes correspond to longer cavity lengths. Bold lines indicate multiples of $L = 40$ nm. Intensity maxima are indicated by red circles. (b-m) Time-frequency spectrograms of SPP WPs corresponding to the bold lines in (a). White arrows indicate positions of intensity holes.

## 4. Discussion

The cavity length-dependent spectral modulations in both the transmitted and reflected SPP WPs were clearly derived from the spectral filtering effect of the MIM nano-cavity: when an incident SPP WP reached a MIM nano-cavity, only those frequency components that coincided with the cavity's resonance were transmitted, whereas the rest were reflected. Because the line widths of resonances are broad as a consequence of the short coherence lifetimes of MIM nano-cavities, the transmitted SPP maintained a broad spectral width to form femtosecond WPs. Therefore, by selecting the cavity length that allows a single resonance mode to remain in the bandwidth of the SPP, the MIM nano-cavity splits the incoming chirped SPP WP into a femtosecond pulse as the transmission wave and a femtosecond dark-pulse as the reflection wave.

To clarify these relations, time-frequency spectrograms of transmitted (Fig. 7e) and reflected (Fig. 8e) SPP WPs obtained with the 160-nm-long nano-cavity are again shown in Fig. 9(b,c), together with the spectrogram of a SPP on a flat surface (Fig. 4b) as the reference, shown in Fig. 9(a). The vertical and horizontal dashed lines in Fig. 9(b,c), respectively, indicate the coordinates of the intensity peak of the transmitted WP (82 fs, 1.58 eV). These values agree well with the coordinates of the "hole" in Fig. 9(c). Moreover, both time-frequency distributions of the transmission and reflection lie inside the distribution of the reference. Figure 9(d) and (e), respectively, show envelopes of the temporal waveform and FFT spectra of the reference, transmission, and reflection WPs. The time and energy of the transmission peak are indicated by dashed lines. Again, the transmitted and reflected SPP WPs show a complementary relation in both the time and energy domains.

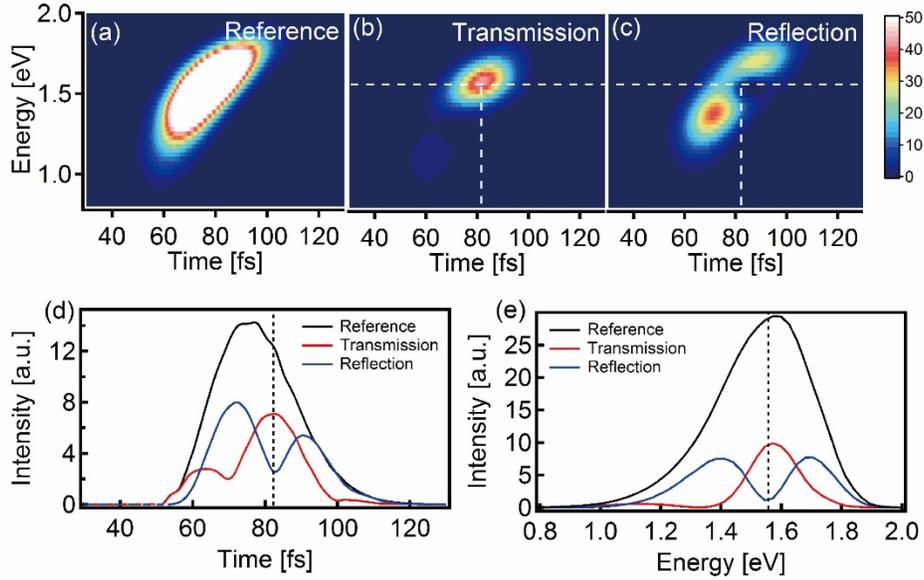

Fig.9. Time-frequency spectrograms of SPP WP prepared from measurements at: (a) $B_T$ of the flat surface as the reference, (b) $B_T$ with a MIM nano-cavity as the transmitted wave, and (c) $B_R$ with a MIM nano-cavity as the reflected wave. The length of the nano-cavity was 160 nm. (d) Envelope shapes of WPs. Black, red, and blue lines are, respectively, the reference wave, the transmitted wave, and the reflected wave. (e) Spectra of the reference wave, the transmitted wave, and the reflected wave. Color assignments are same as in (d). The vertical dashed lines in (b), (c), and (d) indicate the time (82 fs) when the transmitted WP had the maximum intensity. The horizontal dashed lines in (b), (c), and (e) indicate the energy (1.58 eV) of the intensity peak of the transmitted WP.

    The time of the transmission peak is delayed from the reference peak by approximately 5 fs. The sign and the amount of delay were determined by the chirp of the incident SPP WP and the frequency of the cavity resonance. The spectral window of the cavity resonance simply excised a spectral component and let that component through the nano-structure. The amount of delay induced by transmission through an approximately 100-nm scale cavity acting as a waveguide was less than 1 fs. When the incident SPP WP was up-chirped, a cavity resonance set at higher (lower) than the center frequency of the WP induced a positive (negative) delay in the peak of the transmitted WP. In the case studied here, the maximum available delay was approximately ±5 fs.

    Figure 10(a) shows a two-dimensional plot of the FFT spectra of transmitted (red contour) and reflected (blue contour) SPP WPs as a function of the cavity length, $L$. The marked points were taken from Fig. 3(b) to show the 1st–5th order resonance peaks calculated by the FDTD simulation. Overall, the intensities of both the transmission and the reflection oscillated periodically as a function of $L$, reflecting the Fabry-Pérot resonance condition formulated in Eq. (11). The period of oscillation along the horizontal direction corresponded to a half wavelength of an eigenmode of the MIM structure. The transmission and reflection components showed an antiphase relation, providing the complemental spectra observed in Fig. 9(e). Cavity resonances definitely coincided with the transmission peaks and with the reflection dips. The oscillatory behavior of the transmission was less clear than the reflection because of optical interference with diffraction waves. An intensity profile of the transmission component made by slicing the contour at the energy of 1.58 eV (horizontal dashed line in Fig. 10a) is shown in Fig. 10(b) (open circles). The intensity of transmission was well-fitted by a sum of five Gaussian functions (solid line). The peak position of each Gaussian (dashed lines) showed good agreement with the intersection points between the 1st–5th order resonance peaks and the horizontal dashed line at

1.58 eV. This agreement is a favorable indication of the optical functionality of the MIM nano-cavity as a low-finesse Fabry-Pérot etalon.

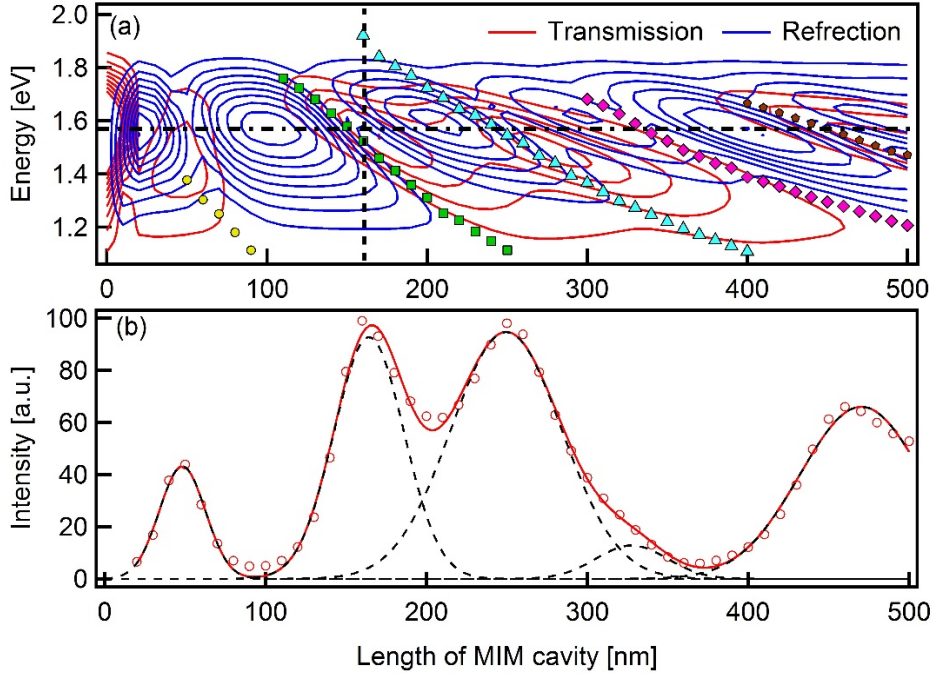

Fig. 10. (a) Two-dimensional plots of FFT spectral intensities of the transmitted (red contour) and reflected (blue contour) WPs shown as a function of the length of the MIM nano-cavity and the energy. Marked points show resonance peaks for the 1st (yellow), 2nd (green), 3rd (blue), 4th (pink), and 5th (brown) order resonance modes of the MIM nano-cavities, as derived from FDTD simulations. The profile sections along the vertical dashed line (length of cavity = 160 nm) is shown in Fig. 9(e). (b) Sectional view of the contour plot of transmitted WPs along the energy of 1.58 eV (horizontal dashed-dotted line in (a)) is shown (red circles). The overall tendency is well-fitted by a numerical sum (solid line) of five Gaussian functions (dashed lines).

The MIM nano-cavities studied here were based on a laminar construction, which is well-suited to today's nano-fabrication techniques, such as nano-lithography and thin film growth. Cavities with lengths of several tens of nanometers can be fabricated using these methods, and some properties of optical filtering could be tuned, such as with thicker insulator layers to narrow transmission widths. If silver is used as the metal material, the resonance frequency can be tuned over the whole visible range. The deep subwavelength dimension enables high-density arrangements of nano-cavities, and together with carrier-wave engineering of ultrafast light fields, these plasmonic elements could enable the manipulation of SPP waves in frequency, temporal, and spatial domains.

## 5. Conclusion

Optical interactions of MIM nano-cavities with femtosecond SPP WPs have been systematically investigated by using FDTD simulation and time-frequency analyses. The SPP WP, including visible to near infrared spectral range, was launched from an excitation source on a continuous Au film, and propagated on the surface to interact with a MIM structure. The incident SPP WP resonantly excited the nanocavity, as well as caused diffraction and scattering, resulted in splitting of the WP to a transmitted and a reflected components remaining on the Au surface as the SPP modes. The resonance nature of the SPP-cavity interaction was reflected in modulations of the spectrum and the envelope shape of the transmitted/reflected WP. Time-frequency spectrograms showed the transmitted WPs possess a few peaks of which energy correspond to eigenmodes of the MIM cavity, while reflected WPs exhibit intensity dips at the corresponding time-frequency coordinates. These features evidence the MIM nano-cavities functioned as Fabry–Pérot etalon-type spectrum filter, which separated WPs into resonance frequencies and others. Because the carrier wave of the SPP WP was chirped, excising a portion of spectral component provoked a production of wavepacket having a narrower time duration and the time of the intensity maximum shifted from that of the reference SPP WP. The amount of peak shift in time was tunable within a range of approximately 10 fs, limited by the amount of chirp, the width of the spectrum of the incident WP, and the widths of resonance peaks and the free spectral ranges of the MIM cavity. These manipulations of SPP WPs in time-frequency domain will expand the controllability of SPP waves in planar dimensions, and will provide useful insights for designing optical devices based on plasmonics and meta-materials.

**Appendices**

*Hole position*

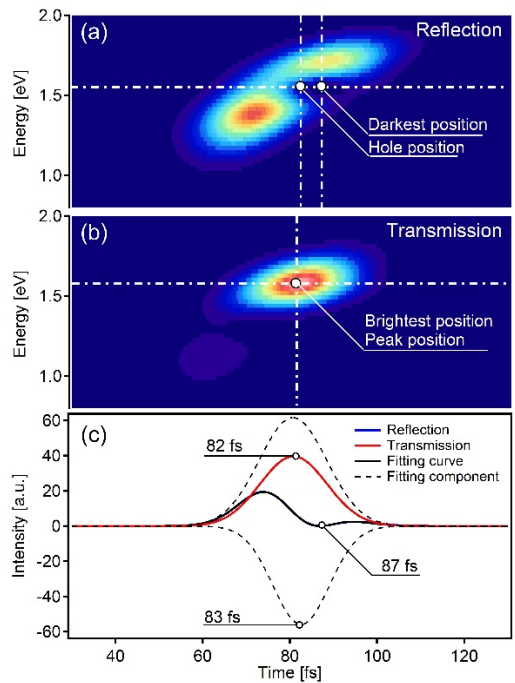

Fig. 11. Spectrograms of (a) reflected wave, and (b) transmitted wave for the length of the cavity of 160 nm. (c) The blue and the red solid lines, respectively, show sectional views of spectrograms of the reflection ($E = 1.55$ eV), and the transmission ($E = 1.58$ eV), indicated by horizontal dashed-dotted lines in (a, b). (The blue line is overlapped by a fitting curve shown as the black solid line.) The reflection curve is well fitted by using two Gaussian functions (dashed lined) each of them represent the positive and the negative intensity components.

In the text, we define the "hole" in the reflected wave packet generated by the 160-nm-long cavity to be located at the time of 83 fs. This value is actually shifted by 4 fs from the time when the deepest dip appears in the spectrogram shown in the Fig. 9(c) (87 fs), and rather close to the time when the transmission peak appears in Fig. 9(b) (82 fs). This discrepancy can be explained as follows. Because the whole shape of the reflected wave packet can be considered as a dark pulse incused in a SPP wave packet having the original waveform, the sectional views of spectrograms would be expressed as a superimposition of a negative pulse and a positive pulse. Fig. 11(c) shows that the sectional view of the reflected wave at $E = 1.55$ eV is reasonably fitted along this line. The least square fitting determined the positions of the two peaks as 82 fs for the positive pulse and 83 fs for the negative pulse. The observed time shift in the intensity dip happened because both the positive and negative pulses had comparable pulse widths. In this paper, we define the "hole position" as the minimum position of the negative pulse.

**Acknowledgements**

The authors would like to thank H. T. Miyazaki for valuable discussions, guidance on MIM structures, and supports for calculations of dispersion curves. This work was supported by JSPS KAKENHI Grant Numbers JP14459290, JP16823280, JP18967972.